\documentclass[journal]{IEEEtran}

\ifCLASSINFOpdf
   \usepackage[pdftex]{graphicx}
    \graphicspath{ {./images/} }
    \DeclareGraphicsExtensions{.pdf,.jpeg,.png, .eps}
\else
\fi
\usepackage{amsmath}
\usepackage{array}
\usepackage{mathtools}
\usepackage{cite}
\usepackage[hidelinks]{hyperref}
\usepackage{graphicx, caption, subcaption}
\usepackage{amsmath,amssymb,amsfonts}
\usepackage{algorithmic}
\usepackage{graphicx}
\usepackage{xcolor}
\usepackage[version=4]{mhchem}
\usepackage{textcomp}
\usepackage{siunitx}
\DeclareSIUnit \turns {turns}

\usepackage[utf8]{inputenc}
\usepackage{float}
\usepackage{ulem}

\begin{document}

\title{Modeling non-planar coils in a full-scale stellarator}
%
%
%

\author{M. Backmeyer, N. Riva, M. Lyly, A. Halbach, and V. Lahtinen
\thanks{M. Backmeyer, M. Lyly, A. Halbach, and V. Lahtinen are with Quanscient Oy (Tampere, Finland) (e-mail:merle.backmeyer@quanscient.com,
 mika.lyly@quanscient.com, alexandre.halbach@quanscient.com, valtteri.lahtinen@quanscient.com)}
\thanks{N. Riva is with MIT Plasma Science and Fusion Center (MA, USA) (e-mail:nicoriva@mit.edu)}
\thanks{This work was supported by Quanscient and by CFS and PSFC-MIT via RPP27. We want to thank Jorrit Lion (jlion@proximafusion.com) for providing the geometry of the stellarator.}}

\markboth{Modeling non-planar coils in a full-scale stellarator}{}

\maketitle

\begin{abstract}
Design and modeling of a stellarator fusion reactor is a multidisciplinary effort that requires a tight integration between simulation of highly nonlinear multi-physics and representation of non-planar complex geometries. The critical current calculation and the design of the mechanical structures are among the most crucial aspects as they set size, cost, and time to build the stellarator. Because of the asymmetric and non-planar nature of its components the modeling of such figures of merit needs to be carried out at large scale, without the possibility of taking advantage of any particular symmetry. In this work we develop a three-dimensional model for the analysis of the magnetic field and forces, necessary for such considerations, for  complex coil geometries, such as stellarators, where a two-dimensional approach can not provide accurate analyses and verification of assumptions. 
Moreover, this method can quickly generate a large amount of critical modeling data (e.g. Lorentz load, displacement and stresses)  that could be integrated into a workflow for coil design optimization based on machine learning or other recent optimization tools.

\end{abstract}


\section{Introduction}
\label{sec:introduction}
\IEEEPARstart{T}{he} most advanced approach for nuclear fusion involves using toroidal magnetic confinement systems, specifically Tokamaks and Stellarators. Both systems use magnetic fields to confine the plasma. In both Tokamaks and Stellarators, particles can drift off course and hit the chamber walls due to the inhomogeneous magnetic field. In Tokamaks one set of coils generates a toroidal field, a central solenoid operated in pulsed mode induces a
toroidal current in the plasma, and a third set of coils generates an outer poloidal field that shapes and positions the plasma. The combination of the field components result in a twisted magnetic field that confines the particles in the plasma. Stellarators confine the plasma by means of non-planar coils only, eliminating the need for the solenoid/transformer. This allows Stellarators to operate without plasma disruption (typically problematic for Tokamaks) and in a steady state for continuous energy production~\cite{pedersen2016confirmation,pedersen2018first}.
For the stellarator coil optimization in the design process, it is crucial to have an accurate simulation of the magnetic field and loads on the coils to meet physics requirements and engineering constraints. 
Therefore, Finite Element Modeling (FEM) tools are mandatory to analyze the electro-magneto-thermal and mechanical response of a high-temperature superconductor (HTS) non-planar coil~\cite{Riva_2023} and predict its safe-operation behavior. Even with an accurate knowledge of the coil geometry, it is challenging to predict its performance accurately. Not only too large Lorentz loads can damage the tape and reduce the coil's performance~\cite{Gates2018}, but excessive deformation of the coil systems can lead to a poor performance of the plasma confinement itself~\cite{Boozer2020}.\\
The non-planarity of stellarator coils, may require to solve full scale three-dimensional (3D) FEM problems with much larger number of degrees of freedom with respect to Tokamaks due to the absence of trivial symmetries (such as the axisymmetric one in tokamaks)~\cite{Gates_2017}. The computational limit with sequential computing or even loop-level parallelism is quickly reached. In this work we present the analysis of the magnetic field and Lorentz loads of a full-scale stellarator based on the Helias-3 configuration~\cite{Helias2001,Helias2004} using the domain-decomposition methods (DDM) which leads to short computation times and high accuracy. This is a continuation of our previous work, where we introduced \texttt{Quanscient Allsolve}, a custom cloud-based simulation tool in the context of DDM-accelerated quench and AC loss simulations \cite{riva2023h}.

Computational times of large and complex FEM scale models in the order of minutes could enable the use of Artificial Intelligence (AI) tools to corroborate and perform generative design to optimize the mechanical structures.

In section \ref{sec:theoryMaterial} we describe the procedure to obtain the coil orientations , briefly mention the formulation and describe the utilized custom DDM tool. In section \ref{sec:results}, we present our results, compared to results using \texttt{COMSOL}, and a mesh sensitivity study. Finally, in section \ref{sec:results} we draw conclusions.

\section{Computation model}
\label{sec:theoryMaterial}
\begin{figure}[tb]
\centering
\includegraphics[width=0.8\linewidth]{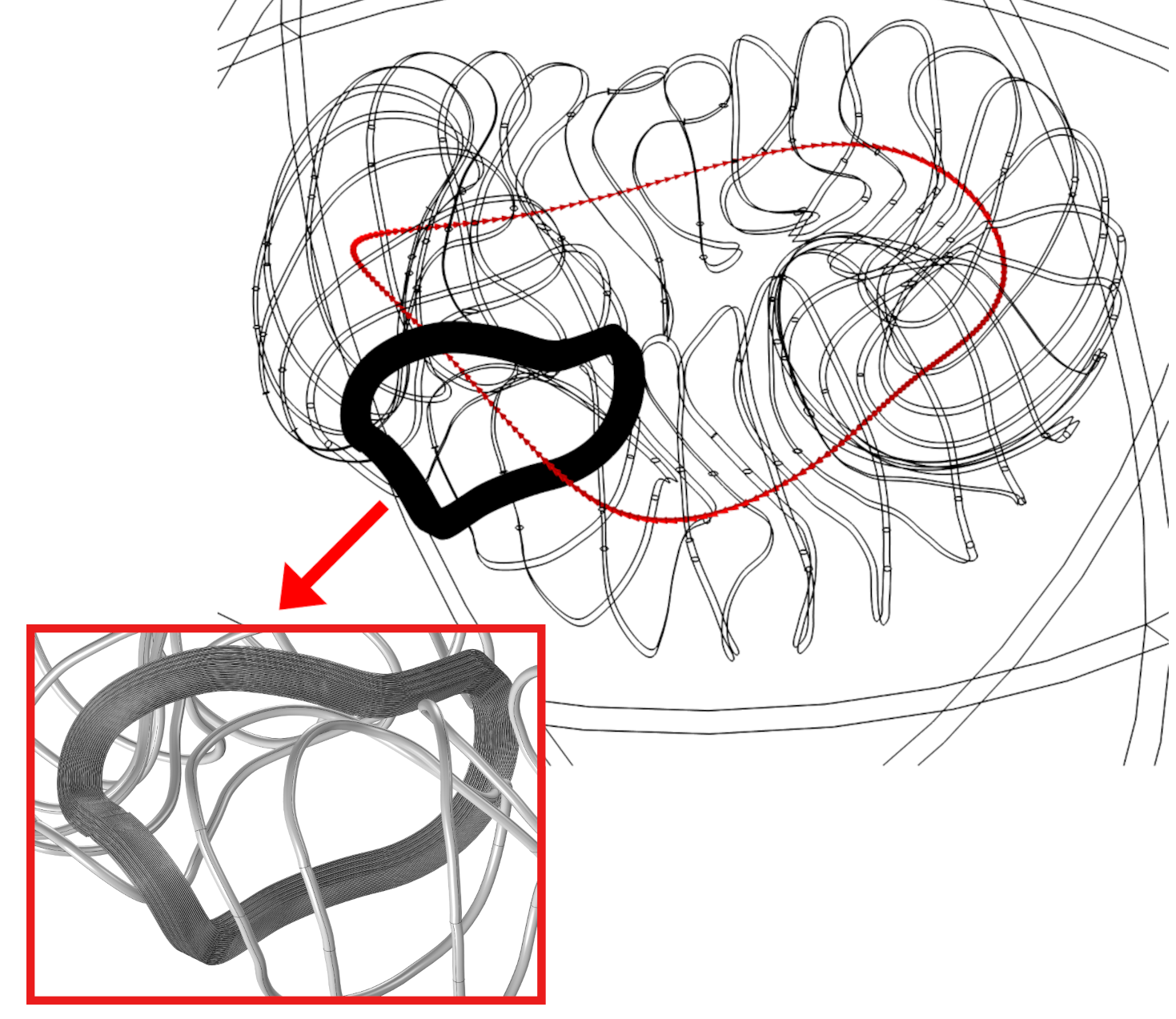}
\caption{The geometry of the full stellarator with a diameter of approx. \SI{20}{\meter}.}
\label{fig:stellarator_geo}
\end{figure}

The stellarator geometry that will be analyzed is shown in Figure~\ref{fig:stellarator_geo} and is based on the HELIAS-3~\cite{Helias2001,Helias2004}. This configuration consists of 30 coils 
and produces an average magnetic field on the magnetic axis (the "middle line" of the plasma in Fig \ref{fig:stellarator_geo}) of \SI{8.9}{\tesla}. The theoretical value is calculated using the Biot-Savart law. One of the purposes of this work is to evaluate the Lorentz loads on the coils and use this as an input (for future works) to design the mechanical radial plates. Simulating in high-detail (full-turns) all the coils is impractical and unnecessary. A possible approach is to focus on the mechanical design of a single coil (See Figure~\ref{fig:stellarator_geo} inset) and simulate the overall magnetic field experienced by that coil (self-field + remaining coils) where $N_{\rm{coil}}-1$ coils are homogenized (all the turns are considered to be homogenized in one single turn) and the single coil of interest is simulated in great detail. 
This approach allows to focus on one coil at the time and generate data (i.e. Lorentz loads) for a reduced and separate model (3D mechanical) that can be used for coil design in an optimization workflow.

Specifically, in this work the stellarator is composed of a total of 30 non-planar coils: 29 coils are homogenized with rectangular cross-section and carry \SI{13.27}{\mega\ampere}; one of the coils only is modeled on a detailed level with 208 turns (see Figure~\ref{fig:stellarator_geo} inset) and carries ${I_{op}}=\SI{63798}{\ampere}$ (i.e. \SI{13.27}{\mega\ampere}/208).
A magnetostatic field map study of the full-stellarator assembly is carried out by applying the current density to each of the 29 homogenized coil as well as to the 208 turns of the detailed coil (see section~\ref{sec:MagA}). Note that given the complex geometry of each coil, it's necessary to determine local curvilinear coordinates aligned with the coil/turn orientations via numerical method similar to the methods utilized in~\cite{VargasLlanos_2022,Riva_2023} to introduce the current density normal to each coil. In these works the utilized software was \texttt{COMSOL}, where a Curvilinear Coordinate module (or the Coil Geometry Analysis) can be used for this purpose. In this work we introduce a new method based on the Laplace equations (see section~\ref{sec:CC}). \\
Finally, solving the problem was accelerated using Domain Decomposition Method (DDM) on an Amazon Web
Services (AWS) cloud infrastructure suited for multiphysics DDM, provided by Quanscient (\url{www.quanscient.com}) under the \texttt{Quanscient Allsolve} software. Its distributed memory management ensures memory requirements will not be a bottleneck.

\subsection{Magnetostatic $A$ Formulation}
\label{sec:MagA}
The computational domain $\Omega$ consists of air $\Omega_{air}$ and conducting coils $\Omega_{coil}$ and is enclosed by the domain boundary $\partial\Omega$.
The well-known $\mathbf{A}$-formulation is solved to find the distribution of the the magnetic vector potential $\mathbf{A}$.
The discretization of the magnetic vector potential $A$  is carried out using Whitney edge elements \cite{bossavit1988whitney}.

\subsection{Curvilinear Coordinates}
\label{sec:CC}
\begin{figure}[t]
\centering
\includegraphics[width=0.7\linewidth,trim=0cm 2cm 0cm 2cm]{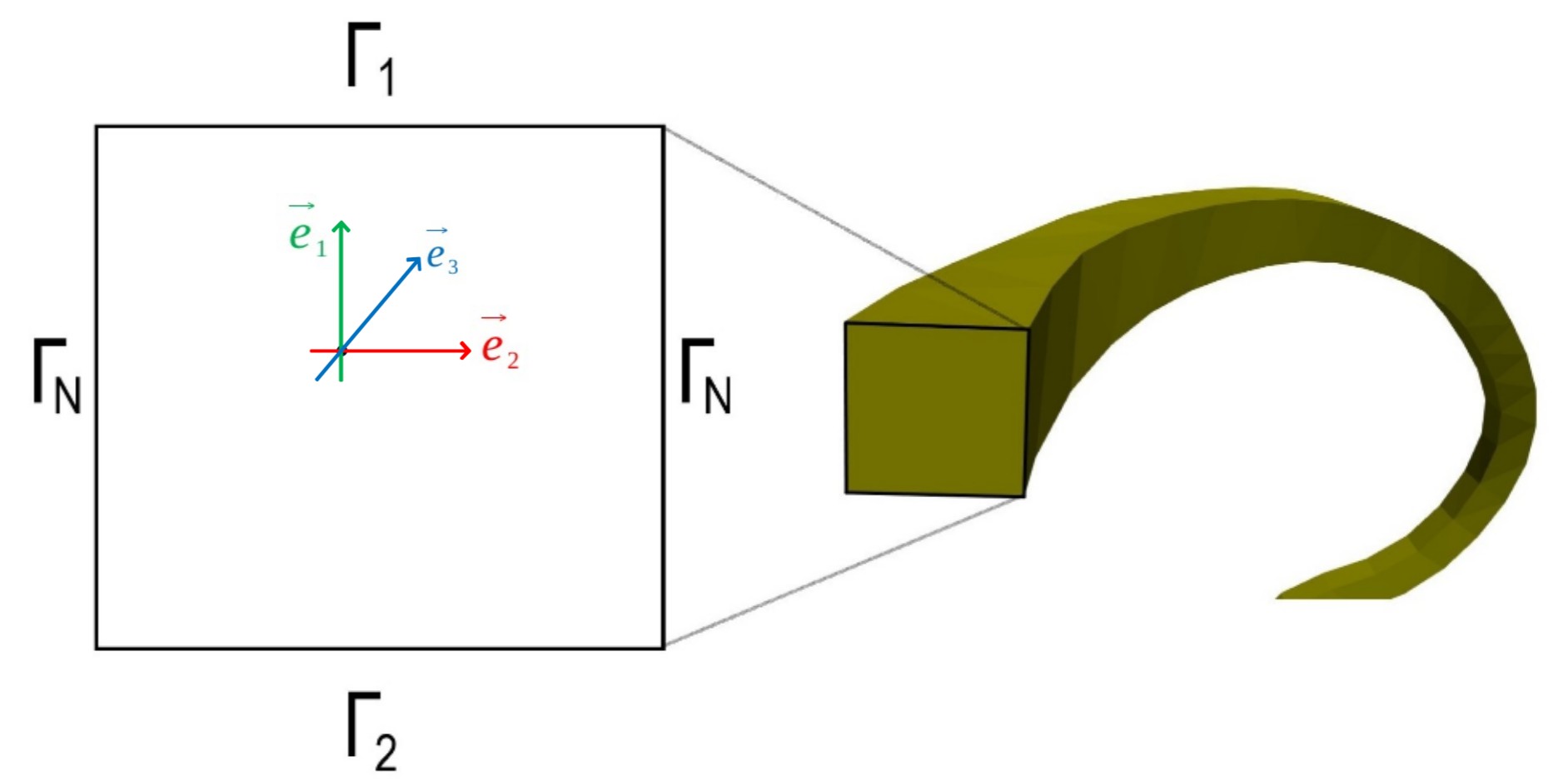}
\caption{Setup for the computation of the cable orientation.}
\label{fig:orientation}
\end{figure}

To introduce the current density $\mathbf{J}$ oriented along the cable, the curvilinear coordinates of the cables/turns need to be determined. A local coordinate system is introduced at each point in the cable (see Figure \ref{fig:orientation}) and described by the gradients of the three scalar potentials $\varphi_1$, $\varphi_2$ and $\varphi_3$, i.e.
\begin{equation}
    \Vec{e}_1 = \nabla \varphi_1, \Vec{e}_2 = \nabla \varphi_2, \Vec{e}_3 = \nabla \varphi_3.
\end{equation}

Exploiting the mutual orthogonality of the surfaces $\Gamma_1$ and $\Gamma_2$, the scalar potential $\varphi_1$ is obtained solving a Laplace problem. It represents the solution to the boundary value problem,
\begin{align}
\Delta \varphi_1 &= 0 \text{ on } \Omega \\
\varphi_1 &= 0 \text{ on } \Gamma_1 \\
\varphi_1 &= 1 \text{ on } \Gamma_2 \\
\mathbf{n} \cdot {\nabla} \varphi_1 &= 0 \text{ on } \Gamma_N.
\end{align}

Repeating this procedure for $\varphi_2$ with interchanged boundary conditions
\begin{align}
\varphi_2 &= \begin{cases} 0\\ 1\end{cases} \text{ on } \Gamma_N \\
\mathbf{n} \cdot {\nabla} \varphi_2 &= 0 \text{ on } \Gamma_1 \text{ and } \Gamma_2,
\end{align}
we get with $\nabla \varphi_1$ and $\nabla \varphi_2$ two base unit vectors of the local curvilinear coordinate system at any point of the coil. The cross product of $\nabla \varphi_1$ and $\nabla \varphi_2$ defines the third basis vector of the coordinate system, which is orthogonal to the cross section and represents the direction of the current flow  $\frac{\mathbf{J}}{\|\mathbf{J}\|}$, so

\begin{equation}
    \nabla \varphi_1 \times \nabla \varphi_2 = \frac{\mathbf{J}}{\|\mathbf{J}\|}.
\end{equation}

\begin{figure}[tb]
\centering
\begin{subfigure}{0.49\linewidth}
    \centering
    \includegraphics[width=\linewidth,trim=0cm 1cm 0cm 1cm,clip]{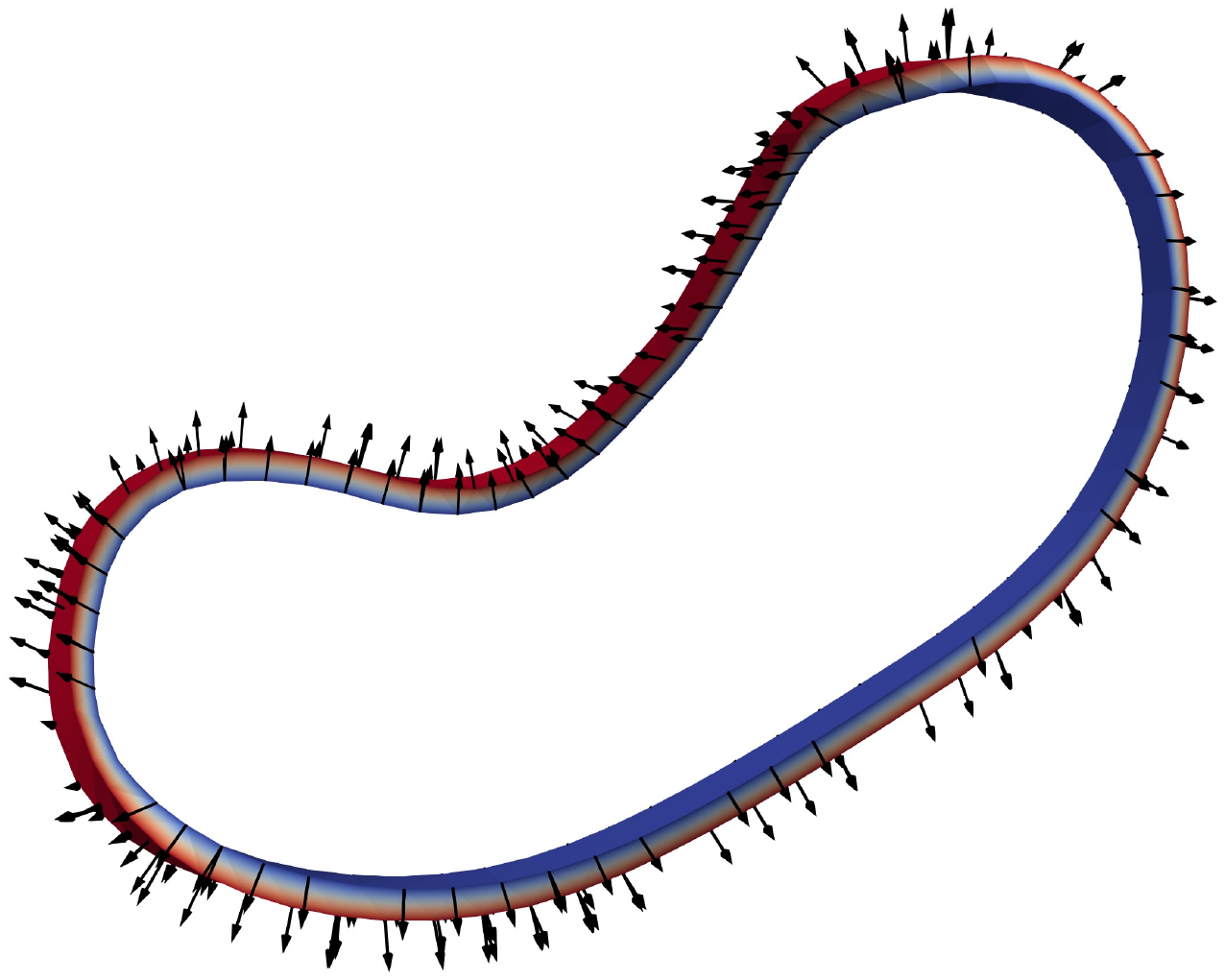}
    \caption{$\varphi_1$ as colorplot and the normal vectors $\nabla \varphi_1 $ as vectors.}
    \label{fig:orient1}
\end{subfigure}
\begin{subfigure}{0.49\linewidth}
    \centering
    \includegraphics[width=\linewidth,trim=0cm 1cm 0cm 1cm,clip]{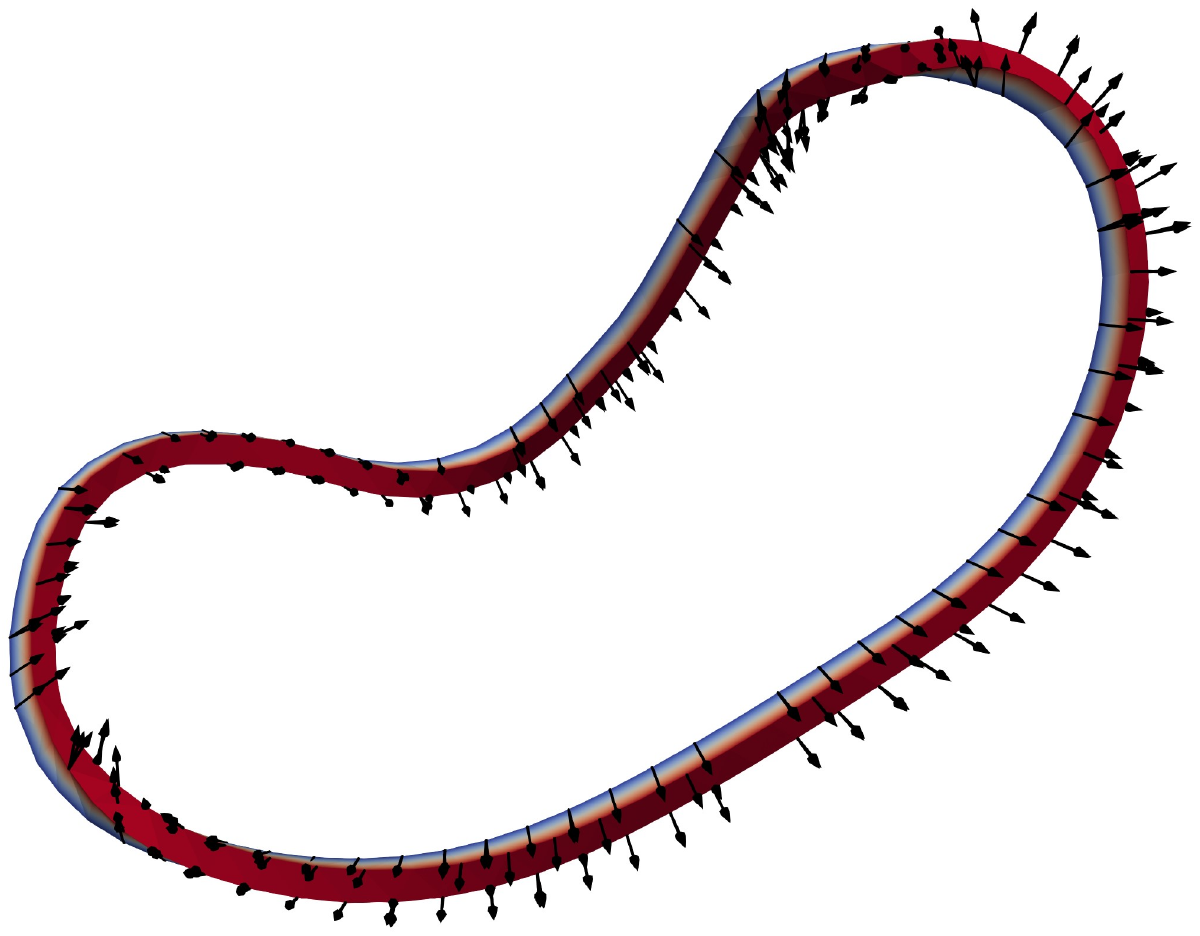}
    \caption{$\varphi_2$ as colorplot and the normal vectors $\nabla \varphi_2$ as vectors.}
    \label{fig:orient2}
\end{subfigure}

\begin{subfigure}{\linewidth}
    \centering
    \includegraphics[totalheight=2in]{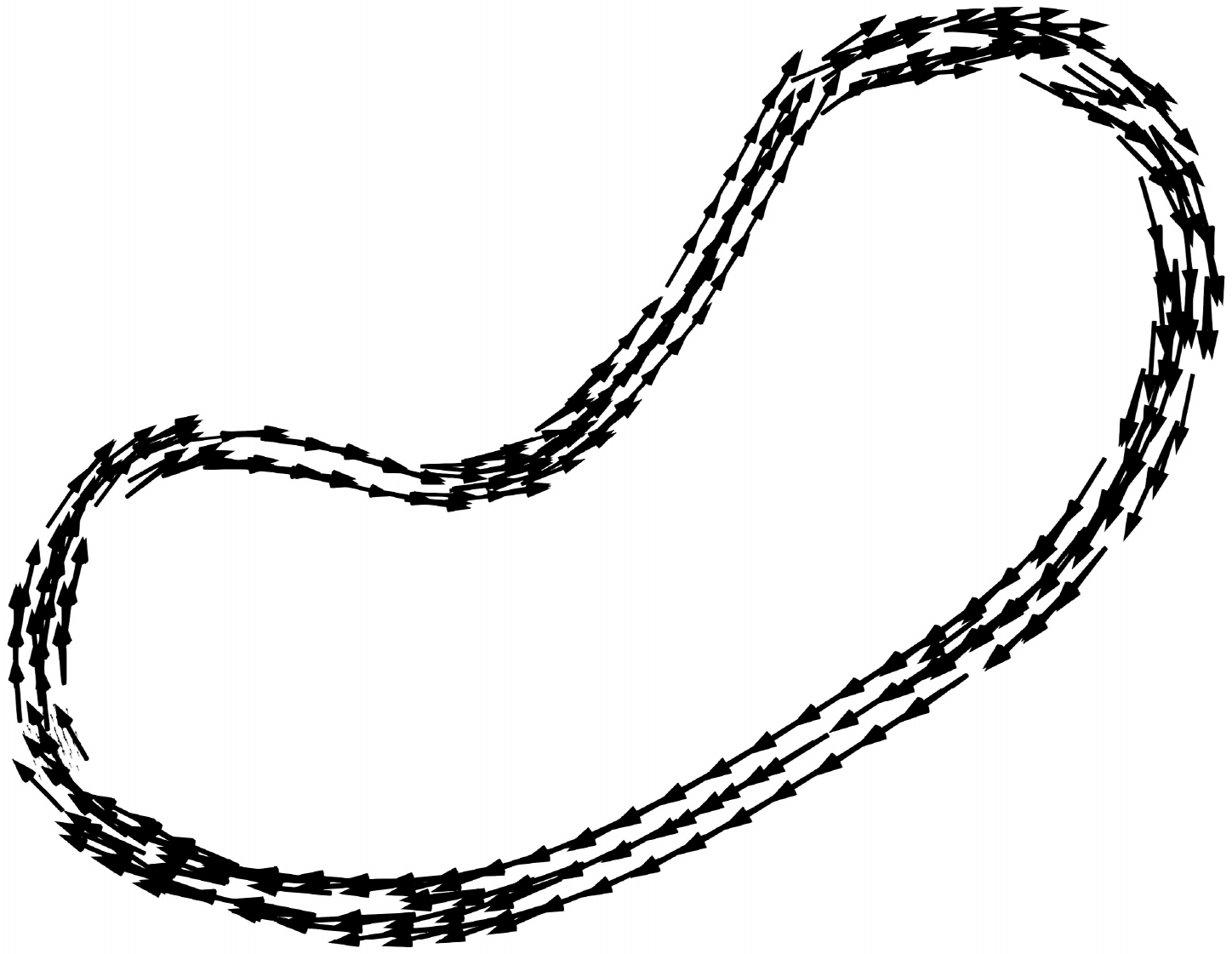}
    \caption{The resulting coil orientation $\frac{\mathbf{J}}{\|\mathbf{J}\|}$ as vector plot.}
    \label{fig:orient3}
\end{subfigure}
\caption{Intermediate steps to find the orientations of the coils/turns.}
\label{fig:orientation_glyphs}
\end{figure}

Figure \ref{fig:orientation_glyphs} shows the intermediate steps of the methods on one exemplary coil. This approach can also be used to model the dependency of the critical current on the magnetic field orientation. In that case, it is used to determine the perpendicular and parallel field components for any tape orientation.

\subsection{Domain Decomposition Method}
\label{sec:DDM}

The domain decomposition method (DDM) in the FEM con-
text refers to the partitioning of the computational mesh into similarly sized pieces that can each be processed on different computing instances, allowing to distribute the computational burden. The DDM method is used to improve the speed of numerical simulations in solid mechanics, electromagnetism, flow in porous media, etc., on parallel machines from tens to hundreds of thousands of cores. This is well suited to take advantage of supercomputer or cloud architectures, which was already shown in \cite{riva2023h}. In this work, the optimized Schwarz algorithm is used in the DDM framework: this algorithm solves the problems defined on the smaller mesh pieces iteratively using generalized minimal residual method (GMRES) and exchanges boundary data between domains at each iteration to reach convergence. Compared to the pioneering work of \cite{schwarz1869ueber} the method used in this work accelerates convergence using GMRES and optimized
boundary data \cite{lions1988schwarz}.

\section{Results}
\label{sec:results}
\subsection{Field plots}
\begin{figure}[tb]
\centering
\includegraphics[width=\linewidth, trim = 5cm 4cm 0cm 5cm, clip]{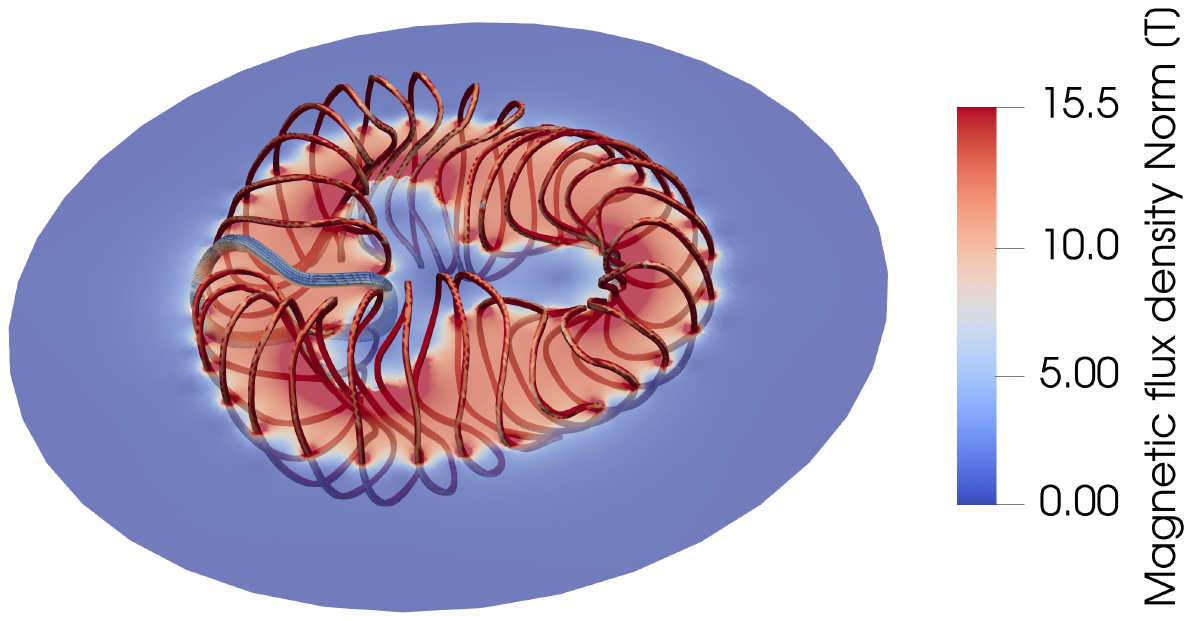}
\caption{Magnetic flux density of the full-scale stellarator.}
\label{fig:mag_field_plot}
\end{figure}

The numerical study is conducted for different interpolation orders and mesh resolutions.
First of all, the resulting fields for the coarser mesh and cubic elements are presented. 
In Figure~\ref{fig:mag_field_plot}, the distribution of the magnetic field in the full stellarator is shown. It matches well the expected theoretical value of $\SI{8.9}{\tesla}$ on the magnetic axis. 

\begin{figure}[tb]
\centering
\includegraphics[width=\linewidth,trim=1cm 4cm 0cm 4cm,clip]{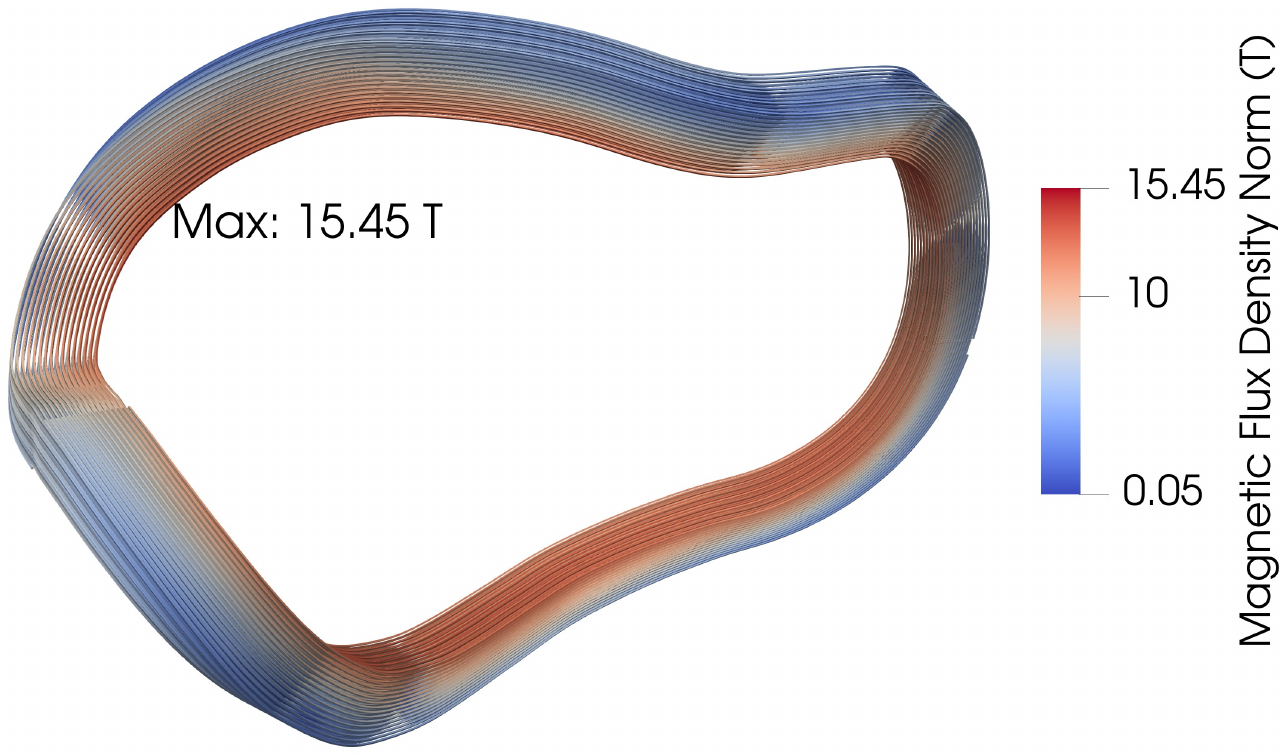}
\caption{Magnetic field $B$ on the full-scale full-turns detailed coil.}
\label{fig:magnetic_field_plot_detail}
\end{figure}

\begin{figure}[tb]
\centering
\includegraphics[width=\linewidth,trim=2cm 3cm 0cm 4cm,clip]{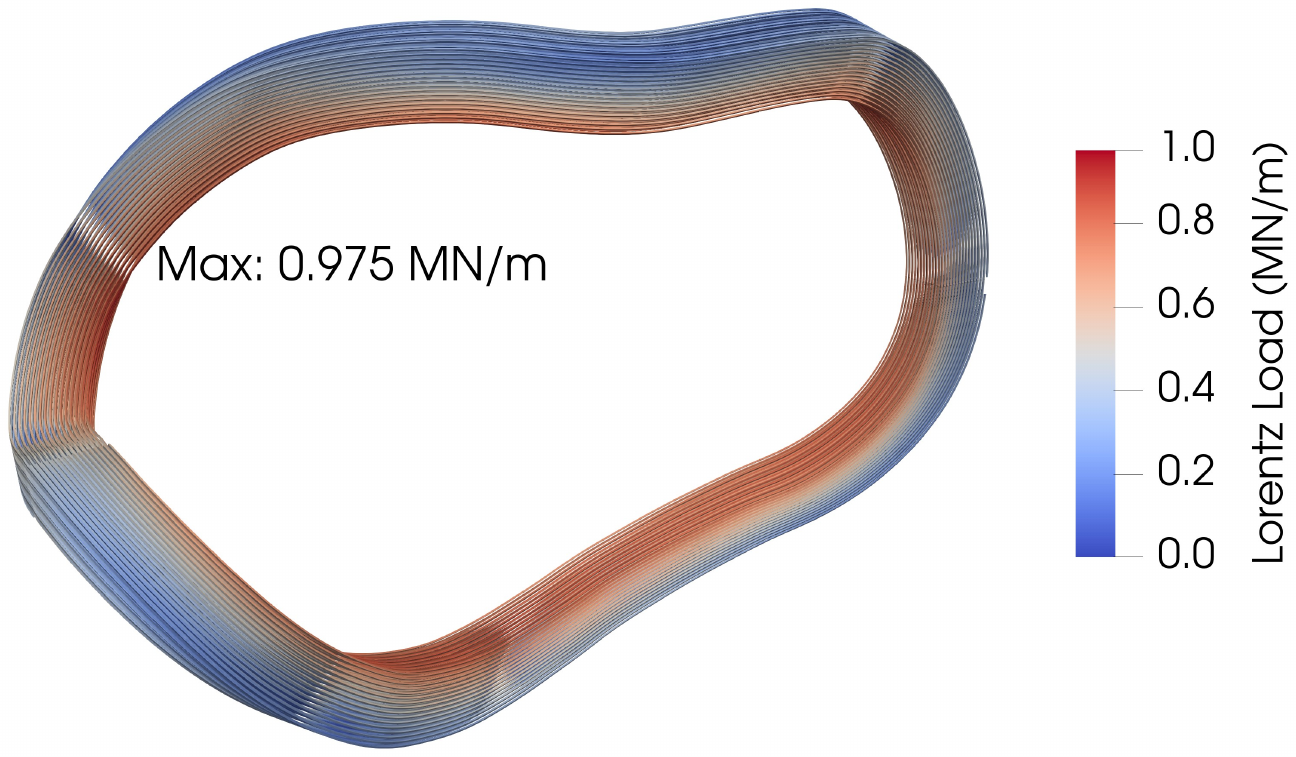}
\caption{Lorentz loads on the full-scale full-turns detailed coil.}
\label{fig:force_field_plot}
\end{figure}

Figure~\ref{fig:magnetic_field_plot_detail} depicts the magnetic field map on the cable volume, and its peak is \SI{15.45}{\tesla}. If we consider  an $I_{\rm{op}}=\SI{63798}{\ampere}$ and a $B_{\rm{peak}}=\SI{15.45}{\tesla}$ we expect a peak of the Lorentz load of $F_{\rm{dens}}=I\times B\simeq\SI{980}{\kilo\newton\per\meter}$. Figure~\ref{fig:force_field_plot} represents the Lorentz force density defined as $I\times B$ that could be used as an input for a mechanical model to evaluate what are the maximum stresses experienced by the coil. The maximum of the Lorentz load of \SI{0.975}{\mega\newton\per\meter} is in reasonable agreement with the rough calculation estimated above and with the maximum Lorentz loads in literature \cite{Hartwig2020,Zhao2022,Zhao_2022_v2,Pierro2020}.

\subsection{Validation against COMSOL Multiphysics}

\begin{figure}[tb]
  \centering
  \includegraphics[width=\linewidth]{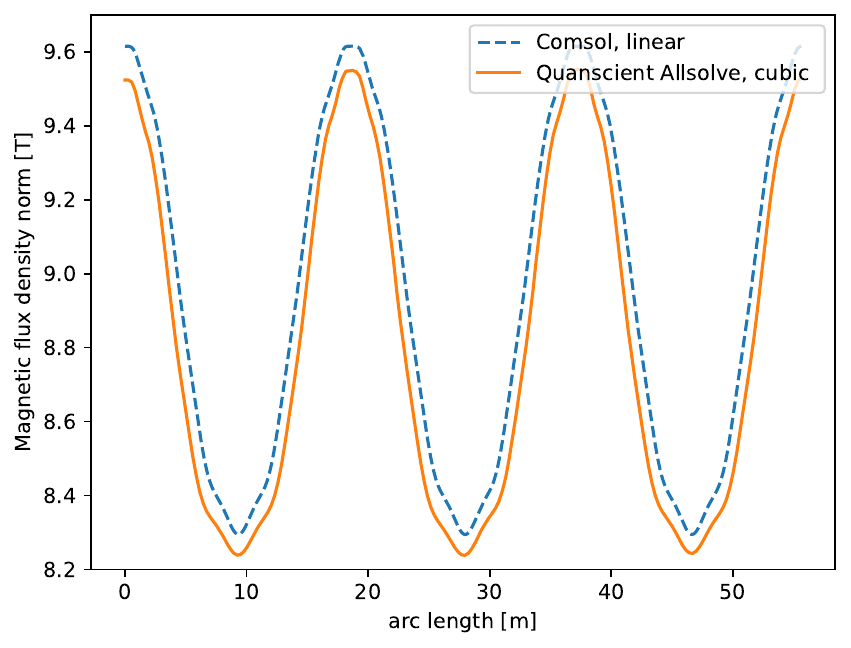}
\caption{Comparison of the magnetic field flux along magnetic axis between \texttt{COMSOL} (dashed) and \texttt{Quanscient.allsolve} (continuous).}
\label{fig:comparison_comsol}
\end{figure}

The solution was validated against  another simulation software, namely the commerical software \texttt{COMSOL Multiphysics}~\cite{COMSOL}, in which a direct solver for the coil analysis, and an iterative solver with Auxiliary-Space Maxwell (AMS) preconditioner for the magnetic problem were used. In both simulation tools, the discretization was performed using linear elements.
Figure~\ref{fig:comparison_comsol} shows the comparison of the magnitude of the magnetic flux density on the magnetic axis (marked in red in Figure~\ref{fig:stellarator_geo}) obtained from both simulations of the stellarator. The different coil orientation acquisition methods (Coil Geometry Analysis vs. Laplace Problem) lead to computations on distinct mesh configurations. The analysis with \texttt{COMSOL} was performed on coils represented with circular cross sections. That explains the slight mismatch of 0.66~\% of the peak values.

The runtime statistics of both simulations will be compared in subsection \ref{subsec:runtime}.




\subsection{Mesh sensitivity analysis}
\begin{figure}[tb]
\centering
\includegraphics[width=\linewidth]{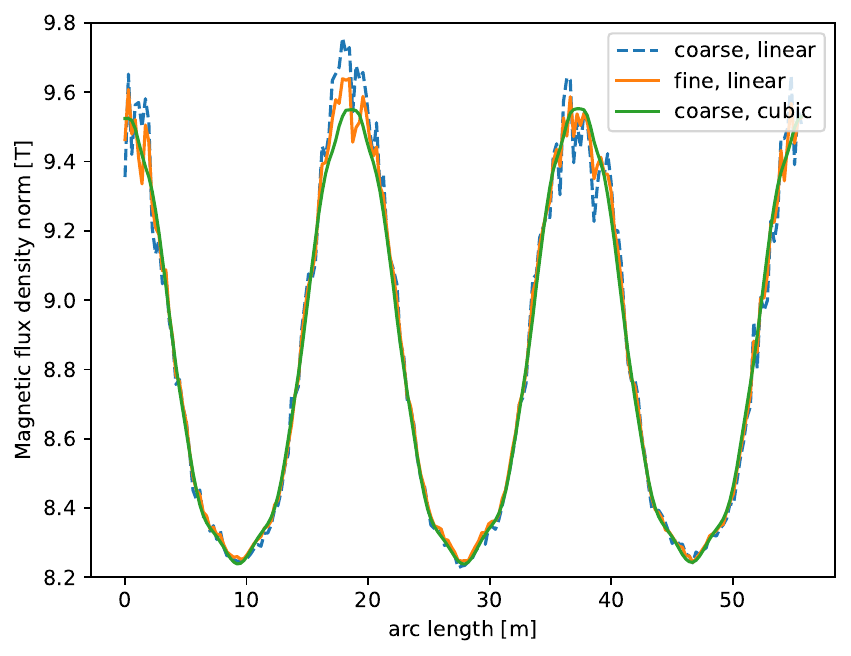}
\caption{Magnetic flux density norm on axis of the stellarator compared for different orders and mesh resolutions.}
\label{fig:comparison_order}
\end{figure}

As a next step, we study the dependency of the solution on the mesh resolution and the interpolation order.
The same simulation was repeated for a finer mesh, obtained by splitting the elements, and for linear and cubic interpolation order. Results are presented in Figure~\ref{fig:comparison_order}.
Increasing the interpolation order leads to smoother field results. The solutions of the two mesh resolutions for cubic interpolation differ by a RMSE below $0.8\%$, indicating that all effects are already covered by the coarser mesh. However, the ability to perform large simulations on different mesh resolutions enables the use of machine learning models for (mechanical) optimization. Multi-level machine learning algorithms learn on the differences of the observable on successive mesh resolutions\cite{lye2021multi}.   

\subsection{Runtime statistics}
\label{subsec:runtime}
Table \ref{tab:runtime} states the number of unknowns, the number of cores used and the computation time for different model setups. 
The advantage of using DDM is clearly visible if we compare the different simulations that were run on \texttt{Quanscient Allsolve}. For example, an increase of the number of DoFs by around $8$ (see column $2$ and $3$) has rarely any effect on the runtime if the number of cores is adapted as well.
Using DDM with \texttt{Quanscient Allsolve} makes it even possible to run the largest problem of over $317$ million DoFs in $17$ minutes.

\begin{table}
  \centering
  \caption{Comparison of the computation times for different setups.}
  \begin{tabular}{p{1.2cm}|p{1.2cm}|p{1.5cm}|p{1.5cm}|p{1.5cm}|p{1.5cm}}
    \hline
    Software & \texttt{COMSOL} & \texttt{Quanscient Allsolve} & \texttt{Quanscient Allsolve} & \texttt{Quanscient Allsolve}\\
    \hline
     Shape function order & linear & linear & cubic & cubic \\
     \hline
     Mesh resolution & coarse & coarse & coarse & fine \\
     \hline
    MDoFs & $1.5$ & $5$ & $39$ & $317$ \\
    \hline
    Computer & $16$ cores, $96$ GB RAM  & $15$ Cores & $50$ Cores & $500$ Cores x $2$ CPUs \\
    \hline
    Runtime (min) & $23$ & $<4$ & $4$ & $17$\\
    \hline
  \end{tabular}
  \label{tab:runtime}
\end{table}

\section{Conclusion}
\label{sec:conclusion}

In this paper, a magnetostatic simulation of a full-scale stellarator was carried out. Utilizing \texttt{Quanscient Allsolve}, an in-house Domain Decomposition Method (DDM) tool was benchmarked against \texttt{COMSOL}, while also demonstrating better scalability on cloud infrastructure. In addition, the Laplace problem was solved to individuate the correct current flow along the complex path of the stellarator coils. 

 A single coil was modeled in high detail, while homogenizing the others. This allowed us to capture the magnetic field and, consequently, Lorentz loads experienced by the chosen coil, considering both its self-field and contributions from the remaining coils.

The results, when compared with those obtained from \texttt{COMSOL}, showcased a high degree of accuracy. Furthermore, a mesh sensitivity analysis showed the advantages of higher interpolation orders over finer mesh refinements.

Ultimately, this method serves as a stepping stone to rapidly generate large amount of critical data such as Lorentz loads, displacement, and stresses. These data can be integrated into a workflow for coil design optimization, potentially leveraging machine learning or other contemporary tools.



\pagebreak

\bibliographystyle{IEEEtran}
\bibliography{Main}

\begin{thebibliography}{10}
\providecommand{\url}[1]{#1}
\csname url@samestyle\endcsname
\providecommand{\newblock}{\relax}
\providecommand{\bibinfo}[2]{#2}
\providecommand{\BIBentrySTDinterwordspacing}{\spaceskip=0pt\relax}
\providecommand{\BIBentryALTinterwordstretchfactor}{4}
\providecommand{\BIBentryALTinterwordspacing}{\spaceskip=\fontdimen2\font plus
\BIBentryALTinterwordstretchfactor\fontdimen3\font minus
  \fontdimen4\font\relax}
\providecommand{\BIBforeignlanguage}[2]{{%
\expandafter\ifx\csname l@#1\endcsname\relax
\typeout{** WARNING: IEEEtran.bst: No hyphenation pattern has been}%
\typeout{** loaded for the language `#1'. Using the pattern for}%
\typeout{** the default language instead.}%
\else
\language=\csname l@#1\endcsname
\fi
#2}}
\providecommand{\BIBdecl}{\relax}
\BIBdecl

\bibitem{pedersen2016confirmation}
T.~S. Pedersen, M.~Otte, S.~Lazerson, P.~Helander, S.~Bozhenkov, C.~Biedermann,
  T.~Klinger, R.~C. Wolf, and H.-S. Bosch, ``Confirmation of the topology of
  the wendelstein 7-x magnetic field to better than 1: 100,000,'' \emph{Nature
  communications}, vol.~7, no.~1, p. 13493, 2016.

\bibitem{pedersen2018first}
T.~S. Pedersen, R.~K{\"o}nig, M.~Krychowiak, M.~Jakubowski, J.~Baldzuhn,
  S.~Bozhenkov, G.~Fuchert, A.~Langenberg, H.~Niemann, D.~Zhang \emph{et~al.},
  ``First results from divertor operation in wendelstein 7-x,'' \emph{Plasma
  Physics and Controlled Fusion}, vol.~61, no.~1, p. 014035, 2018.

\bibitem{Riva_2023}
\BIBentryALTinterwordspacing
N.~Riva, R.~S. Granetz, R.~Vieira, A.~Hubbard, A.~T. Pfeiffer, P.~Harris,
  C.~Chamberlain, Z.~S. Hartwig, A.~Watterson, D.~Anderson, and R.~Volberg,
  ``Development of the first non-planar rebco stellarator coil using viper
  cable,'' \emph{Superconductor Science and Technology}, vol.~36, no.~10, p.
  105001, aug 2023. [Online]. Available:
  \url{https://dx.doi.org/10.1088/1361-6668/aced9d}
\BIBentrySTDinterwordspacing

\bibitem{Gates2018}
D.~A. Gates, D.~Anderson, S.~Anderson, M.~Zarnstorff, H.~Weitzner, D.~Ruzic,
  D.~Andruczyk, A.~Reiman, J.~D. Lore, M.~Landreman, J.~P. Freidberg, S.~R.
  Hudson, M.~Porkolab, D.~Demers, J.~Terry, E.~Edlund, S.~A. Lazerson,
  N.~Pablant, R.~Fonck, F.~Volpe, J.~Canik, R.~Granetz, A.~Ware, J.~D. Hanson,
  S.~Kumar, C.~Deng, K.~Likin, A.~Cerfon, A.~Ram, A.~Hassam, S.~Prager,
  C.~Paz-Soldan, M.~J. Pueschel, I.~Joseph, and A.~H. Glasser,
  \emph{{Stellarator Research Opportunities: A Report of the National
  Stellarator Coordinating Committee}}.\hskip 1em plus 0.5em minus 0.4em\relax
  Springer US, 2018, vol.~37, no.~1.

\bibitem{Boozer2020}
A.~H. Boozer, ``{Why carbon dioxide makes stellarators so important},''
  \emph{Nuclear Fusion}, vol.~60, no.~6, 2020.

\bibitem{Gates_2017}
\BIBentryALTinterwordspacing
D.~Gates, A.~Boozer, T.~Brown, J.~Breslau, D.~Curreli, M.~Landreman,
  S.~Lazerson, J.~Lore, H.~Mynick, G.~Neilson, N.~Pomphrey, P.~Xanthopoulos,
  and A.~Zolfaghari, ``Recent advances in stellarator optimization,''
  \emph{Nuclear Fusion}, vol.~57, no.~12, p. 126064, oct 2017. [Online].
  Available: \url{https://dx.doi.org/10.1088/1741-4326/aa8ba0}
\BIBentrySTDinterwordspacing

\bibitem{Helias2001}
\BIBentryALTinterwordspacing
C.~Beidler, E.~Harmeyer, F.~Herrnegger, Y.~Igitkhanov, A.~Kendl, J.~Kisslinger,
  Y.~Kolesnichenko, V.~Lutsenko, C.~Nührenberg, I.~Sidorenko, E.~Strumberger,
  H.~Wobig, and Y.~Yakovenko, ``The helias reactor hsr4/18,'' \emph{Nuclear
  Fusion}, vol.~41, no.~12, p. 1759, dec 2001. [Online]. Available:
  \url{https://dx.doi.org/10.1088/0029-5515/41/12/303}
\BIBentrySTDinterwordspacing

\bibitem{Helias2004}
T.~Andreeva, C.~D. Beidler, E.~Harmeyer, Y.~L. Igitkhanov, Y.~I. Kolesnichenko,
  V.~V. Lutsenko, A.~Shishkin, F.~Herrnegger, J.~Kißlinger, and H.~F.~G.
  Wobig, ``The helias reactor concept: Comparative analysis of different field
  period configurations,'' \emph{Fusion Science and Technology}, vol.~46,
  no.~2, pp. 395--400, 2004.

\bibitem{riva2023h}
N.~Riva, A.~Halbach, M.~Lyly, C.~Messe, J.~Ruuskanen, and V.~Lahtinen, ``{$ H
  $}-{$\phi$ }formulation in sparselizard combined with domain decomposition
  methods for modeling superconducting tapes, stacks, and twisted wires,''
  \emph{IEEE Transactions on Applied Superconductivity}, vol.~33, no.~5, pp.
  1--5, 2023.

\bibitem{VargasLlanos_2022}
C.~R. Vargas-Llanos, F.~Huber, N.~Riva, M.~Zhang, and F.~Grilli, ``3d
  homogenization of the t-a formulation for the analysis of coils with complex
  geometries,'' \emph{Superconductor Science and Technology}, vol.~35, no.~12,
  p. 124001, oct 2022.

\bibitem{bossavit1988whitney}
A.~Bossavit, ``Whitney forms: A class of finite elements for three-dimensional
  computations in electromagnetism,'' \emph{IEE Proceedings A (Physical
  Science, Measurement and Instrumentation, Management and Education,
  Reviews)}, vol. 135, no.~8, pp. 493--500, 1988.

\bibitem{schwarz1869ueber}
\BIBentryALTinterwordspacing
H.~Schwarz, ``Ueber einige abbildungsaufgaben.'' vol. 1869, no.~70, pp.
  105--120, 1869. [Online]. Available:
  \url{https://doi.org/10.1515/crll.1869.70.105}
\BIBentrySTDinterwordspacing

\bibitem{lions1988schwarz}
P.-L. Lions \emph{et~al.}, ``On the schwarz alternating method. i,'' in
  \emph{First international symposium on domain decomposition methods for
  partial differential equations}, vol.~1.\hskip 1em plus 0.5em minus
  0.4em\relax Paris, France, 1988, pp. 1--42.

\bibitem{Hartwig2020}
Z.~S. Hartwig, R.~Vieira, B.~N. Sorbom, R.~A. Badcock, M.~Bajko, W.~K. Beck,
  B.~Castaldo, C.~L. Craighill, M.~Davies, J.~Estrada, V.~Fry,
  T.~Golfinopoulos, A.~E. Hubbard, J.~H. Irby, S.~Kuznetsov, C.~J. Lammi,
  P.~Michael, T.~Mouratidis, R.~A. Murray, A.~T. Pfeiffer, S.~Z. Pierson,
  A.~Radovinsky, M.~D. Rowell, E.~E. Salazar, M.~Segal, P.~W. Stahle,
  M.~Takayasu, T.~L. Toland, and L.~Zhou, ``{VIPER: An industrially scalable
  high-current high temperature superconductor cable},'' \emph{Superconductor
  Science and Technology}, no.~33, p. 11LT01, 2020.

\bibitem{Zhao2022}
Z.~Zhao, P.~Moore, and L.~Chiesa, ``Structural modeling of rebco viper cable
  for high-field magnet applications,'' \emph{IEEE Transactions on Applied
  Superconductivity}, vol.~32, no.~6, pp. 1--5, 2022.

\bibitem{Zhao_2022_v2}
------, ``Structural finite element analysis of rebco tape delamination with
  solid-shell element under various loads,'' \emph{IOP Conference Series:
  Materials Science and Engineering}, vol. 1241, no.~1, p. 012032, may 2022.

\bibitem{Pierro2020}
F.~Pierro, Z.~Zhao, L.~Chiesa, G.~Celentano, M.~Marchetti, and A.~della Corte,
  ``Structural modeling of hts cable-in-conduit conductor with helically
  slotted aluminum core for high-field magnet applications,'' \emph{IEEE
  Transactions on Applied Superconductivity}, vol.~30, no.~4, pp. 1--5, 2020.

\bibitem{COMSOL}
\BIBentryALTinterwordspacing
``{COMSOL Multiphysics{\textregistered} v. 6.0. (Website), COMSOL AB,
  Stockholm, Sweden}.'' [Online]. Available: \url{www.comsol.com}
\BIBentrySTDinterwordspacing

\bibitem{lye2021multi}
K.~O. Lye, S.~Mishra, and R.~Molinaro, ``A multi-level procedure for enhancing
  accuracy of machine learning algorithms,'' \emph{European Journal of Applied
  Mathematics}, vol.~32, no.~3, pp. 436--469, 2021.

\end{thebibliography}

\end{document}